\documentclass[10pt,conference]{IEEEtran}

\usepackage{amsmath}
\newtheorem{theorem}{\bf Theorem}
\newtheorem{proposition}{\bf Proposition}
\newtheorem{conjecture}{\bf Conjecture}

\newtheorem{corollary}{\bf Corollary}
\newtheorem{definition}{\bf Definition}

\newcommand{\EE}{{\mathcal{E}}}

\begin{document}
\title{Concavity of entropy under thinning}

\author{
\IEEEauthorblockN{Yaming Yu}
\IEEEauthorblockA{Department of Statistics\\
University of California\\
Irvine, CA 92697, USA\\
Email: yamingy@uci.edu}
\and
\IEEEauthorblockN{Oliver Johnson}
\IEEEauthorblockA{Department of Mathematics\\
University of Bristol\\
University Walk, Bristol, BS8 1TW, UK\\
Email: o.johnson@bristol.ac.uk}
}
\maketitle
\begin{abstract}
Building on the recent work of Johnson (2007) and Yu (2008), we prove that entropy is
a concave function  with respect to the thinning operation $T_\alpha$. 
That is, if $X$ and $Y$ are independent random variables on $\mathbf{Z}_+$ with ultra-log-concave 
probability mass functions, then
$$H(T_\alpha X+T_{1-\alpha} Y)\geq \alpha H(X)+(1-\alpha)H(Y),\quad 0\leq \alpha\leq 1,$$
where $H$ denotes the discrete entropy.  This is a discrete analogue of the inequality ($h$ denotes the differential entropy) 
$$h(\sqrt{\alpha} X + \sqrt{1-\alpha} Y)\geq \alpha h(X)+(1-\alpha) h(Y),\quad 0\leq \alpha\leq 1,$$
which holds for continuous $X$ and $Y$ with finite variances
and is equivalent to Shannon's entropy power inequality.
  As a consequence we establish a special case of a conjecture of Shepp and Olkin (1981).  
 Possible extensions are also discussed. 
\end{abstract}

\section{Introduction}
This paper considers information-theoretic properties of the thinning map,
an operation on the space of discrete random variables, based on random summation. 
\begin{definition}[R\'{e}nyi, \cite{R}]
For a discrete random variable $X$ on $\mathbf{Z}_+=\{0, 1, \ldots\}$, the thinning operation $T_\alpha$ is defined by 
$$T_\alpha X=\sum_{i=1}^X B_i$$
where $B_i$ are (i) independent of each other and of $X$ and (ii) identically distributed Bernoulli$(\alpha)$ random variables, i.e., $\Pr(B_i=1)=1-\Pr(B_i=0)=\alpha$ for each
$i$.
\end{definition}

Equivalently, if the probability mass function (pmf) of $X$ is $f$, then the pmf of $T_\alpha X$ is 
$$(T_\alpha f)_i\equiv \Pr(T_\alpha X=i)=\sum_{j\geq i} bi(i; j, \alpha) f_j,$$ 
where $bi(i; j, \alpha)=\binom{j}{i} \alpha^i(1-\alpha)^{j-i}$ is the binomial pmf.  
(Note that we 
write $T_\alpha$ for the map acting on the pmf as well as acting on 
the random variable.)  

We briefly mention other notation used in this paper. 
We use ${\rm Po}(\lambda)$ to denote the Poisson distribution with mean $\lambda$, i.e., the pmf is $po(\lambda)=\{po(i; \lambda),\ i=0, 1, \ldots\},\ po(i;\lambda)=\lambda^i e^{-\lambda}/i!$.  The entropy of a discrete random variable $X$ with pmf $f$ is defined as 
$$H(X)=H(f)=\sum_i -f_i\log f_i,$$ 
and the relative entropy between $X$ (with pmf $f$) and $Y$ (with pmf $g$) is defined as 
$$D(X||Y)=D(f||g)=\sum_i f_i\log (f_i/g_i).$$
For convenience we write $D(X)=D(X||po(\lambda))$
where $\lambda=EX$. 

The thinning operation is intimately associated with the Poisson distribution and Poisson convergence theorems.  
It plays a significant role in the derivation of a maximum entropy property for the Poisson distribution (Johnson \cite{J07}). Recently there has been evidence
that, in a number of problems related to information theory,
the operation $T_\alpha$ is the discrete counterpart 
of the operation of scaling a random variable by $\sqrt{\alpha}$; see \cite{HJK, HJK2, J07, Y08}.  Since scaling arguments can give simple proofs of results such as the Entropy Power Inequality, we believe that improved understanding
of the thinning operation could lead to discrete analogues of such results.

For example, thinning lies at the heart of
the following result (see \cite{HJK, HJK2, Y08}), which is a Poisson limit theorem with an information-theoretic interpretation. 
\begin{theorem}[Law of Thin Numbers]
\label{thm1}
Let $f$ be a pmf on $\mathbf{Z}_+$ with mean $\lambda<\infty$.  Denote by $f^{*n}$ the $n$th convolution of $f$, i.e., the pmf of $\sum_{i=1}^n X_i$ where $X_i$ are independent and identically distributed (i.i.d.) with pmf $f$.  Then
\begin{enumerate}
\item
$T_{1/n}(f^{*n})$ converges point-wise to ${\rm Po}(\lambda)$ as $n\to\infty$;
\item
$H(T_{1/n} (f^{*n}))$ tends to $H(po(\lambda))$ as $n\to\infty$;
\item
as $n\to\infty$, $D(T_{1/n} (f^{*n}))$ monotonically decreases to zero, if it is ever finite;
\item
if $f$ is ultra-log-concave, then $H(T_{1/n} (f^{*n}))$ increases in $n$.
\end{enumerate}
\end{theorem}

For Part (4), we recall that a random variable $X$ on $\mathbf{Z}_+$ is called {\it ultra-log-concave}, or ULC, if its pmf $f$ is such that the sequence $i! f_i,\ i=0, 1, \ldots,$ is log-concave.  Examples of ULC random variables include the binomial and the Poisson.  In general, a sum of independent (but not necessarily identically distributed) Bernoulli random variables is ULC.  Informally, a ULC random variable is less ``spread out'' than a Poisson with the same mean. Note that in Part (4) the ULC assumption is natural since, among ULC distributions with a fixed mean, the Poisson achieves maximum entropy (\cite{J07, Y08}). 

Parts (2) and (3) of Theorem \ref{thm1} (see \cite{HJK, HJK2}) 
resemble the entropic central limit theorem of Barron \cite{B}, in that convergence in relative entropy, rather than the usual weak convergence, is established.  The monotonicity statements in Parts (3) and (4), proved in \cite{Y08}, can be seen as the discrete analogue of the
monotonicity of entropy in the central limit theorem, conjectured by Shannon and proved
much later by Artstein et al. \cite{Art}.

In this work we further explore the behavior of entropy under thinning.  Our main result is the following concavity property. 

\begin{theorem}
\label{thm2}
If $X$ and $Y$ are independent random variables on $\mathbf{Z}_+$ with ultra-log-concave pmfs, then
\begin{equation}
\label{tineq}
H(T_\alpha X+T_\beta Y)\geq \alpha H(X)+\beta H(Y),\quad \alpha,\, \beta\geq 0,\ \alpha+\beta\leq 1.
\end{equation}
\end{theorem} 

Theorem \ref{thm2} is interesting on two accounts.  Firstly, it can be seen as an analogue of the inequality
\begin{equation}
\label{cont}
h(\sqrt{\alpha} X+\sqrt{1-\alpha} Y)\geq \alpha h(X)+(1-\alpha) h(Y)
\end{equation}
where $X$ and $Y$ are continuous random variables with finite variances and $h$ denotes the differential entropy.  The difference between thinning by $\alpha$ in (\ref{tineq}) and scaling by $\sqrt{\alpha}$ in (\ref{cont}) is required to control different moments. In the discrete case, the law of small numbers \cite{HJK} and the corresponding maximum entropy property \cite{J07} both require control of the mean, which is achieved by this thinning factor. In the continuous case,
the central limit theorem \cite{B} requires control of the variance, which is achieved by this choice of scaling.  It is well-known that (\ref{cont}) is a reformulation of Shannon's entropy power inequality (\cite{S, Bl}).  Thus Theorem \ref{thm2} may be regarded as a first step towards a discrete entropy power inequality (see Section IV for further discussion). 

Secondly, Theorem \ref{thm2} is closely related to an open problem of Shepp and Olkin \cite{SO} concerning Bernoulli sums.  With a slight abuse of notation let $H(a_1, \ldots, a_n)$ denote the entropy of the sum $\sum_{i=1}^n X_i$, where $X_i$ is an independent Bernoulli random variable with parameter $a_i,\ i=1, \ldots, n$. 

\begin{conjecture}[\cite{SO}]
\label{conj}
The function $H(a_1, \ldots, a_n)$ is concave in $(a_1, \ldots, a_n)$, i.e., 
\begin{eqnarray}
\label{shepp}
\lefteqn{H\left(\alpha a_1+ (1-\alpha) b_1, \ldots, \alpha a_n+ (1-\alpha) b_n \right)}
\nonumber \\
& \geq &
\alpha H(a_1, \ldots, a_n)+ 
(1-\alpha) H(b_1, \ldots, b_n)
\end{eqnarray}
for all $0 \leq \alpha \leq 1$ and $a_i, b_i\in [0,1]$.
\end{conjecture}

As noted by Shepp and Olkin \cite{SO}, $H(a_1, \ldots, a_n)$ is concave 
in each $a_i$ and is concave
in the special case where $a_1 = \ldots = a_n$ and $b_1 = \ldots = b_n$.  We provide further evidence supporting Conjecture \ref{conj},  by
proving another special case, which is a consequence of Theorem \ref{thm2} when applied to Bernoulli sums.

\begin{corollary}
\label{coro}
Relation (\ref{shepp}) holds if $a_ib_i=0$ for all $i$.
\end{corollary}

Conjecture \ref{conj} remains open.  We are hopeful, however, that the techniques introduced here could help resolve this long-standing problem. 

In Section II we collect some basic properties of thinning and ULC distributions, which are used in the proof of Theorem \ref{thm2} in Section III.  Possible extensions are discussed in Section IV.

\section{Preliminary observations}
Basic properties of thinning include the semigroup relation (\cite{J07}) 
\begin{equation}
\label{semigroup}
T_\alpha (T_\beta f)=T_{\alpha\beta} f
\end{equation}
and the commuting relation ($*$ denotes convolution)
\begin{equation}
\label{commute}
T_\alpha (f*g)=(T_\alpha f)*(T_\alpha g).
\end{equation}
It is (\ref{commute}) that allows us to deduce Corollary \ref{coro} from Theorem \ref{thm2} easily.

Concerning the ULC property, three important observations (\cite{J07}) are 
\begin{enumerate}
\item
a pmf $f$ is ULC if and only if the ratio $(i+1)f_{i+1}/f_i$ is a decreasing function of $i$;
\item
if $f$ is ULC, then so is $T_\alpha f$; 
\item
if $f$ and $g$ are ULC, then so is their convolution $f*g$.
\end{enumerate}

A key tool for deriving Theorem \ref{thm2} and related results (\cite{J07, Y1}) is Chebyshev's rearrangement theorem, which states that the covariance of two increasing functions of the same random variable is non-negative.  In other words, if $X$ is a scalar random variable, and $g$ and $\tilde{g}$ are increasing functions, then (assuming the expectations
are finite)
$$E[g(X)\tilde{g}(X)]\geq Eg(X) E\tilde{g}(X).$$

\section{Proof of Theorem \ref{thm2}}
The basic idea is to use the decomposition
$$H(X)=-D(X)-L(X)$$
where as before $D(X)=D(X||po(\lambda))$ with $\lambda=EX$, and $L(X)=E\log (po(X; \lambda))$. 

The behavior of the relative entropy $D(X)$ under thinning is fairly well-understood.  In particular, by differentiating $D(T_\alpha X)$ with respect to $\alpha$ and then using a data-processing argument, Yu \cite{Y08} shows that 
\begin{equation}
\label{dthin}
D(T_\alpha X)\leq \alpha D(X).  
\end{equation}
Further, for any independent $U$ and $V$, the data-processing inequality shows that $D(U+V) \leq D(U) + D(V).$
By taking $U = T_\alpha X$ and $V = T_{1-\alpha} Y$, one concludes that 
\begin{align*}
D(T_\alpha X + T_{1-\alpha} Y) &\leq D(T_\alpha X) +  D(T_{1-\alpha} Y) \\
&\leq \alpha D(X) + (1-\alpha) D(Y).
\end{align*}

Therefore we only need to prove the corresponding result for $L$, that is 
\begin{equation}
\label{lineq}
L(T_\alpha X + T_{1-\alpha} Y) \leq \alpha L(X) + (1-\alpha) L(Y).
\end{equation}
Unfortunately, matters are more complicated because there is no
equivalent of the data-processing inequality, i.e., the inequality $L(U+V) \leq L(U) + L(V)$ 
does not always hold. (Consider for example $U$ and $V$ i.i.d.\ Bernoulli
with parameter $p\in (0,1)$.  This inequality then reduces to $2 p \leq p^2$, which 
clearly fails for all $p$.) 

Nevertheless, it is possible to establish (\ref{lineq}) directly.  We illustrate the strategy 
with a related but simpler result, which involves the equivalent of Equation (\ref{dthin}) for $L$.

\begin{proposition}
\label{prop1}
For any pmf $f$ on $\mathbf{Z}_+$ with mean $\lambda<\infty$, we have $H(T_{\alpha} f)\geq \alpha H(f)$.
\end{proposition}
\begin{IEEEproof}
Let us assume that the support of $f$ is finite; the general case follows by a truncation argument (\cite{Y08}).  In view of (\ref{dthin}), we only need to show $l(\alpha)\leq \alpha l(1)$, where 
$$l(\alpha)= L(T_\alpha f) = 
\sum_{i\geq 0} (T_\alpha f)_i \log \left(po(i; \alpha\lambda)\right).$$
By substituting $f(\alpha) = 0$ in
Equation (8) of \cite{J07}, we obtain that
$$\frac{{\rm d} (T_\alpha f)_i}{{\rm d} \alpha} = \frac{i (T_\alpha f)_i - (i+1) (T_\alpha f)_{i+1}}{\alpha},$$
and hence, using summation by parts,
\begin{align*}
l'(\alpha) &= \lambda \log(\alpha \lambda) -
\sum_{i\geq 0} \frac{{\rm d} (T_\alpha f)_i}{{\rm d} \alpha} 
 \log  i! \\
&= \lambda \log(\alpha \lambda) - \frac{1}{\alpha} \sum_{i\geq 0} (i+1) (T_\alpha f)_{i+1}
 \log \left( i+1 \right).
\end{align*}
In a similar way, using the inequality $\log(1+u) \leq u,\ u>-1$, 
\begin{align*}
l''(\alpha) &=\frac{\lambda}{\alpha}- \frac{1}{\alpha^2} \sum_{i\geq 0} (T_\alpha f)_{i+2} (i+2)(i+1)
\log 
\frac{i+2}{i+1}\\
            &\geq \frac{\lambda}{\alpha}- \frac{1}{\alpha^2} \sum_{i\geq 0} (T_\alpha f)_{i+2} (i+2)(i+1)
\frac{1}{i+1}\\
            &= \frac{\lambda}{\alpha}- \frac{1}{\alpha^2} \sum_{i \geq 0} (T_\alpha f)_{i+2} (i+2)
\geq 0.
\end{align*}
The last inequality holds since $\sum_{s=0}^\infty s (T_\alpha f)_s=\lambda \alpha$.  

Having established the convexity of $l(\alpha)$, we can now deduce the full Proposition
using (\ref{dthin}).
\end{IEEEproof} 

Before proving Theorem \ref{thm2}, we note that although (\ref{tineq}) is stated for $\alpha+\beta\leq 1$, only the case $\alpha+\beta=1$ need to be considered.  Indeed, if (\ref{tineq}) holds for $\alpha+\beta=1$, then for general $\alpha,\ \beta\geq 0$ such that $\alpha+\beta=\gamma\leq 1$, we have 
\begin{align}
\nonumber
H(T_{\alpha} X+T_{\beta} Y) &=H(T_\gamma (T_{\alpha/\gamma} X+T_{\beta/\gamma} Y))\\
\label{ineq:prop1}
                          &\geq \gamma H(T_{\alpha/\gamma} X+T_{\beta/\gamma} Y)\\
                          \nonumber
                          &\geq \alpha H(X)+ \beta H(Y),
\end{align}
where (\ref{semigroup}) and (\ref{commute}) are used in the equality, and Proposition \ref{prop1} is used in (\ref{ineq:prop1}). 

\begin{IEEEproof}[Proof of Theorem \ref{thm2}]
Assume $\beta=1-\alpha$, and let $f$ and $g$ denote the pmfs of $X$ and $Y$ respectively.  Assume $\lambda=EX>0$ and $\mu=EY>0$ to avoid the trivial case.  As noted before, we only need to show that 
$$l(\alpha)=\sum_{i\geq 0} (T_\alpha f*T_\beta g)_i\log po(i; \alpha\lambda+\beta\mu)$$
is convex in $\alpha$ (where $\beta=1-\alpha$).  The calculations are similar to (but more involved than) those for Proposition \ref{prop1}, and we omit the details.  The key is to express $l''(\alpha)$ in the following form suitable for applying Chebyshev's rearrangement theorem.
\begin{align*}
l''(\alpha) 
=& \frac{(\lambda-\mu)^2}{\alpha\lambda+\beta\mu}+A+B
\end{align*}
where 
$$A = \sum_{i\geq 1, j\geq 0} (T_\alpha f)_i (T_\beta g)_j i a(i,j),$$
$$B = \sum_{i\geq 0, j\geq 1} (T_\alpha f)_i (T_\beta g)_j j b(i,j),$$
and 
\begin{align*}
a(i,j)  &=
\left( \frac{i+j-1}{\alpha^2} - \frac{\beta \mu j}{(\alpha \lambda + \beta \mu) \alpha^2\beta^2} \right) \log\frac{i+j-1}{i+j}, \\
b(i,j) &= \left(
\frac{i+j-1}{\beta^2} - \frac{\alpha \lambda i}{(\alpha \lambda + \beta \mu)\alpha^2\beta^2} \right) 
\log\frac{i+j-1}{i+j}. \end{align*}
Ultra-log-concavity and dominated convergence permit differentiating term-by-term.  

For each fixed $j$, since $(i+j-1)\log((i+j-1)/(i+j))$ decreases in $i$ 
and $\log((i+j-1)/(i+j))$ increases in $i$, we know that $a(i,j)$ decreases in $i$.  
Since $T_\alpha f$ is ULC, the ratio $i 
(T_\alpha f)_{i}/(T_\alpha f)_{i-1}$ is decreasing in $i$. 
Hence we may apply Chebyshev's rearrangement theorem to the sum over $i$ and obtain 
\begin{align}
\nonumber
A & =
\sum_{i\geq 1, j\geq 0} (T_\alpha f)_{i-1} (T_\beta g)_j \left( \frac{i 
(T_\alpha f)_{i}}{(T_\alpha f)_{i-1}} \right) a(i,j) \\
& \geq \alpha \lambda \sum_{i\geq 1, j\geq 0} (T_\alpha f)_{i-1} (T_\beta g)_j a(i,j) 
\nonumber \\
& = \alpha \lambda \sum_{i, j\geq 0} (T_\alpha f)_{i} (T_\beta g)_j a(i+1,j).
\label{ineq1}
\end{align}
Similarly, considering the sum over $j$, since $b(i,j)$ is decreasing in $j$ for
any fixed $i$,
\begin{align}
\label{ineq2}
B &\geq 
\beta \mu \sum_{i, j\geq 0} (T_\alpha f)_{i} (T_\beta g)_j b(i,j+1).
\end{align}
Adding up (\ref{ineq1}) and (\ref{ineq2}), and noting that 
$$ \alpha \lambda a(i+1,j) + \beta \mu b(i,j+1)
= \frac{(\lambda-\mu)^2}{\alpha\lambda+\beta\mu} (i+j)\log\frac{i+j}{i+j+1},
$$
 we get
\begin{align*}
l''(\alpha)\geq &\frac{(\lambda-\mu)^2}{\alpha\lambda+\beta\mu} \\
& + \sum_{i, j\geq 0} (T_\alpha f)_i (T_\beta g)_j \frac{(\lambda-\mu)^2}{\alpha\lambda+\beta\mu} (i+j)\log\frac{i+j}{i+j+1},
\end{align*}
which is nonnegative, in view of the inequality $u\log(u/(u+1))\geq -1,\ u\geq 0$.
\end{IEEEproof}

\section{Towards a discrete Entropy Power Inequality}
In the continuous case, (\ref{cont}) is quickly shown (see \cite{D})
to be equivalent to Shannon's entropy power inequality
\begin{equation}
\label{entpower}
\exp(2h(X+Y)) \geq \exp(2h(X)) + \exp(2h(Y)),
\end{equation}
valid for independent $X$ and $Y$ with finite variances, with equality if and only
if $X$ and $Y$ are normal.  We aim to formulate a discrete analogue of (\ref{entpower}),
with the Poisson distribution playing the same role as the normal since it has
the corresponding infinite divisibility and maximum entropy properties.

Observe that the function $\exp(2t)$ appearing in (\ref{entpower})
is (proportional to) the inverse of the
entropy of the normal with variance $t$. That is, if we write $e(t) = h(N(0,t))
= \log(\sqrt{2 \pi t})$
then the entropy power 
$v(X) = e^{-1}(h(X)) = \exp(2h(X))/(2 \pi)$, so Equation (\ref{entpower})
can be written as 
$$v(X +Y) \geq v(X) + v(Y).$$

Although there does not exist a corresponding closed form expression for the entropy
of a Poisson random variable, we can
denote ${\cal E}(t)=H(po(t))$.  Then ${\cal E}(t)$ is increasing and concave.  (The proof of Proposition \ref{prop1}, when specialized to the Poisson case, implies this concavity.)  Define $$V(X)={\cal E}^{-1}(H(X)).$$ 
That is, $H(po(V(X)))=H(X)$.  It is tempting to conjecture that the natural discrete analogue of Equation (\ref{entpower}) is  
$$ V(X+Y) \geq V(X) + V(Y),$$
for independent discrete random variables $X$ and $Y$,
with equality if and only if $X$ and $Y$ are Poisson.
However, this is not true.  A counterexample, provided
by an anonymous referee, is the case where $X$ and $Y$ 
both have the pmf
$p(0) = 1/6$, $p(1) = 2/3$, $p(2) = 1/6.$
Since this pmf even lies within the ULC
class, the conjecture still fails when restricted to this class.

We believe that the discrete counterpart of the entropy power inequality
should involve the thinning operation described above. If so, the 
natural conjecture is the following, which we refer to as the thinned
entropy power inequality.

\begin{conjecture}
If $X$ and $Y$ are independent random variables with ULC pmfs on $\mathbf{Z}_+$, then ($0<\alpha<1$)
\begin{equation}
\label{conj2}
V(T_\alpha X+T_{1-\alpha} Y)\geq \alpha V(X)+(1-\alpha) V(Y).
\end{equation}
\end{conjecture}

In a similar way to the continuous case, (\ref{conj2}) easily yields the concavity
of entropy, Equation (\ref{tineq}), as a corollary.  Indeed, by (\ref{conj2}) and the concavity of ${\cal E}(t)$, we have
\begin{align*}
H(T_\alpha X+T_{1-\alpha} Y) &\geq {\cal E}(\alpha V(X)+(1-\alpha) V(Y))\\
                             &\geq \alpha {\cal E} (V(X)) +(1-\alpha) {\cal E} (V(Y))\\
                             &=\alpha H(X)+(1-\alpha) H(Y)
\end{align*}
and (\ref{tineq}) follows.  

Unlike the continuous case, (\ref{tineq}) does not easily yield (\ref{conj2}). The
key issue is the question of scaling. That is, in the continuous case, the
entropy power $v(X)$ satisfies $v( \sqrt{\alpha} X) = \alpha v(X)$ for all $\alpha$
and $X$. It is this result that allows Dembo et al. \cite{D} to deduce
(\ref{entpower}) from (\ref{cont}).

Such an identity does not hold for thinned random variables. However, we conjecture
that \begin{equation} \label{rtepi}
V(T_\alpha X) \geq \alpha V(X) \end{equation} for all $\alpha$ and ULC $X$. 
Note that this Equation (\ref{rtepi}), which we refer to as the restricted thinned
entropy power inequality (RTEPI), is simply the case $Y = 0$ of the full thinned entropy
power inequality (\ref{conj2}). If (\ref{rtepi}) holds, we can use the argument
provided by \cite{D} to deduce the following result, which is in some sense close
to the full thinned entropy power inequality, although $\beta + \gamma < 1$ in general.
\begin{proposition}
\label{prop2}
Consider independent ULC random variables $X$ and $Y$. 
For any $\beta,\ \gamma\in (0,1)$ such that 
$$ \frac{\beta}{1-\gamma} \leq \frac{V(Y)}{V(X)} \leq
\frac{1-\beta}{\gamma},$$
if the RTEPI (\ref{rtepi}) holds then
$$ V(T_\beta X + T_{\gamma} Y) \geq \beta V(X) + \gamma
V(Y).$$
\end{proposition} 
\begin{IEEEproof} Note that an equivalent formulation of the RTEPI
(\ref{rtepi})
is that if $X'$ is Poisson with $H(X) = H(X')$ then for any
$\alpha \in (0,1)$, $H(T_\alpha X) \geq H(T_\alpha X').$
Given $X$ and $Y$ we define $X'$ and $Y'$ to be Poisson with
$H(X) = H(X')$ and $H(X) = H(Y')$.  

Given $\beta$ and $\gamma$, we pick $\alpha$ such that
$\beta \leq \alpha$ and $\gamma \leq 1-\alpha$ so that:
\begin{eqnarray}
\lefteqn{
H(T_\beta X + T_{\gamma} Y) } \nonumber \\ & = & 
H(T_\alpha (T_{\beta/\alpha} X) + T_{1-\alpha} (T_{\gamma/(1-\alpha)} Y)) \nonumber \\
& \geq & \alpha H(T_{\beta/\alpha} X) + (1-\alpha) H (T_{\gamma/(1-\alpha)} Y) \label{eq:pci} \\
& \geq & \alpha H(T_{\beta/\alpha} X') + (1-\alpha) H (T_{\gamma/(1-\alpha)} Y') \label{eq:rtepi} \\
& = & \alpha \EE( \beta V(X)/\alpha) + (1-\alpha) \EE( \gamma V(Y)/
(1-\alpha)) \nonumber
\end{eqnarray}
where Equation (\ref{eq:pci}) follows by Theorem \ref{thm2}
and Equation (\ref{eq:rtepi}) follows by the reformulated RTEPI.

Now making the (optimal) choice 
$$\alpha = \beta V(X)/(\beta V(X) + \gamma V(Y))$$ 
this inequality becomes 
$$H(T_\beta X + T_{\gamma} Y) \geq \EE( \beta V(X) + \gamma V(Y)).$$
The result follows by applying $\EE^{-1}$ to both sides.  Note that the restrictions on $\beta$ and $\gamma$ are required to ensure $\beta \leq \alpha$ and $\gamma \leq 1-\alpha$.
\end{IEEEproof}

Again assuming (\ref{rtepi}), Proposition \ref{prop2} yields the following special case of (\ref{conj2}).  The reason
this argument works is that, as in \cite{D}, if $X$ is Poisson then (\ref{rtepi}) holds with equality for all $\alpha$.

\begin{corollary}
If RTEPI (\ref{rtepi}) holds then (\ref{conj2}) holds in the special case where $X$ is ULC and 
$Y$ is Poisson with mean $\mu$ such that $\mu\leq V(X)$. 
\end{corollary}
\begin{IEEEproof}
For $\gamma\in (0,1)$ let $Z$ be Poisson with mean $\mu(1-\alpha)/\gamma$.  Then 
$V(Z)=\mu(1-\alpha)/\gamma$.  The condition $\mu\leq V(X)$ ensures that we can choose 
$\gamma$ small enough such that 
$$\frac{\alpha}{1-\gamma}\leq \frac{V(Z)}{V(X)}\leq \frac{1-\alpha}{\gamma}.$$
By Proposition \ref{prop2}, 
$$V(T_\alpha X+T_\gamma Z)\geq \alpha V(X)+\gamma V(Z).$$
The claim follows by noting that $T_\gamma Z$ has the same Poisson distribution as $T_{1-\alpha} Y$. 
\end{IEEEproof}

We hope to report progress on (\ref{conj2}) in future work.  Given the fundamental importance of (\ref{entpower}), it would also be interesting to see potential applications of (\ref{conj2}) (if true) and (\ref{tineq}). 
For example, Oohama \cite{O} used the entropy power inequality (\ref{entpower}) to solve
the multi-terminal source coding problem. This showed the rate at which information could be
transmitted from $L$ sources, producing correlated Gaussian signals but unable to collaborate or
communicate with each other, under the addition of Gaussian noise. 
It would be of interest to know whether (\ref{conj2}) could lead to a corresponding
result for discrete channels.

{\bf Note}: Since the submission of this paper to ISIT09, we have found a proof of the restricted thinned 
entropy power inequality, i.e., Equation (\ref{rtepi}).  The proof, based on \cite{J07}, is somewhat technical
and will be presented in a future work.

\end{document}